# Inhibiting Amyloid-like Aggregation through Bio-conjugation of Proteins with Polymer Surfactant

*Anasua Mukhopadhyay, [†,‡] Iliya D. Stoev, [#,‡] David A. King, [#] Kamendra P. Sharma[†,]\* and Erika Eiser [#,]\**

[†] Department of Chemistry, Indian Institute of Technology Bombay, Mumbai 400076, India

[#] Cavendish Laboratory, University of Cambridge, Cambridge CB3 0HE, United Kingdom

Prevention of protein aggregation and thus stabilization of proteins has large biological and biotechnological implications. Here, we show that inhibition of amyloid-like aggregates is possible in stoichiometric conjugates of polymer surfactant and bovine serum albumin (BSA) chosen as a model protein. We investigate using a combination of Thioflavin-T fluorescence spectroscopy, dynamic light scattering and FTIR spectroscopy the aggregation behavior in polymer surfactant modified and unmodified (native) BSA solutions. The BSA-polymer surfactant conjugates are stable up to 5 days under aggregation conditions, while native BSA forms amyloid fibrillar structures. Further, DLS-based micro-rheology studies performed with heat-treated 100 to 200 μM native BSA aggregates provided understanding of the equilibrium elastic (G') and viscous (G") moduli over a very large frequency range, reaching MHz, which are inaccessible using bulk rheology. Our results indicate that after 6 days of aggregation conditions, *G'* showed values between ~ 1.2 to 3.6 Pa corresponding to an entanglement length ($\xi$) of ~ 105 nm. Interestingly, heating 200 μM native BSA solution at 65 ºC for 2 days in a plastic Eppendorf resulted in self-standing films. These films exhibited strong ThT-fluorescence intensity and a predominant β-sheet secondary structure from the FTIR studies, suggesting that self-standing microstructure resulted from hierarchical self-assembly of amyloid fibrils.

**INTRODUCTION**

Misfolding of proteins and their aggregation into amyloid fibrils is known to be the major cause of various neurodegenerative diseases.[1–5] Such amyloid fibrils typically form due to the hydrophobic interactions between the β-sheets of individual proteins and are generally insoluble.[6,7] It was found that the formation of amyloid fibrils is independent of the native structure of the protein, but rather depends on a given amino acid sequence.[8–10] The conditions triggering the aggregation process are still not fully understood. Hence, in the past decades much research focused on preventing proteins from self-assembling into aggregates, because of its huge implications in various medical conditions and preventive medication. Examples are finding causes of various neurodegenerative diseases, preventing aggregation of proteins in the pharmaceutical industry, and also engineering stable biomaterials.[11–14] Therapeutic proteins are showing significant importance in the biopharmaceutical industry, but a major challenge is to avoid formation of aggregates and amyloid-like structures during all stages of the manufacturing process, which can render the protein-based drugs ineffective.[15–18] Side-chain chemical modifications, protein sequence modifications, and covalent attachment of polymeric macromolecules, such as the biocompatible poly(ethylene glycol) (PEG), were some of the approaches to improve the protein stability.[6,19–21] Binding proteins to synthetic molecules non-covalently is another approach to stabilize proteins. The most common additives currently are osmolytes, such as arginine and polyols, which are added in large amounts exceeding the protein-to-additive ratio.[22–25] Addition of polymers such as poly(N-isopropylacrylamide) or poly(propylene oxide), can also stabilize enzymes.[26,27] It was also reported that poly(4-styrene sulfonate) and poly(vinyl sulfate) can inhibit aggregation of denatured Cytochrome C through Coulomb interactions.[28] The factors involved in the prevention of protein aggregation through polymers are mainly achieved through minimizing hydrophobic interactions, strengthening hydrogen bonds, steric interactions and osmotic effects arising from the release of counterions.[29–35] Polyelectrolytes can either enhance or decrease the stability of proteins via Coulomb interactions.[36,37]

Bioconjugation of surface-modified proteins with PEG-based polymer surfactants, leading to water-less protein liquids that freeze below ~ 28 °C, has been shown recently.[38–40] Using a similar approach of bioconjugation, we have shown that the water-less system comprising globular protein bovine serum albumin (BSA) retains its near-native α-helical secondary structure even at temperatures of ~ 100 °C.[41] Here, we report experimental evidence that aqueous solutions of such bioconjugated BSA proteins are stable against heat-induced denaturation and subsequent amyloid-fibril formation. We present a comparative study of aqueous solutions of heat-set native BSA (nBSA), which is known to form fibrils at different conditions (pH and high temperatures[42–46]) and bioconjugates of polymer-surfactant modified BSA (PSpBSA; pBSA: positively charged protein) stabilized via electrostatically linked polymer surfactants. In particular, we used fluorescence spectroscopy, dynamic light scattering (DLS) and FTIR spectroscopic experiments to study the structural changes in native and bioconjugated BSA under fibril formation conditions. Also, we show that DLS-based microrheology enables gaining new insights into the rheological response of protein fibrils in solution. Finally, we report that under certain preparation conditions, film rather than fibril formation of the native BSA is observed.

**EXPERIMENTAL SECTION**

**Materials:** Bovine serum albumin (lyophilized powder, ≥ 96%, A2153), Thioflavin-T (ThT), sodium phosphate monobasic, sodium phosphate dibasic, sodium chloride and glycolic acid ethoxylate lauryl ether were purchased from Sigma-Aldrich. Milli-Q water with a resistivity of ~18.2 MΩ·cm at 25 °C was used for the preparation of all solutions and buffers used in this study. The PBS (saline phosphate buffer, 10 mM, pH = 7.0) was freshly prepared before all aggregation experiments and the pH of the solution was maintained by adding 1 M HCl or 1 M NaOH, as deemed necessary. The pH of the solutions was prepared with a precision of ± 0.01.

**Synthesis of bioconjugates of BSA (PSpBSA):** Protein-polymer surfactant bioconjugates of BSA were prepared in an aqueous solution, involving a three-step process, following the experimental procedure we reported earlier.[41] In short, N,N'-dimethyl-1,3-propanediamine (DMAPA; 2 M; pH 6.5) was covalently linked to the surface-accessible acidic residues of the nBSA by carbodiimide (EDC) mediated activation method, followed by electrostatic coupling of the anionic polymer surfactant (PS) using glycolic acid ethoxylate lauryl ether.

**Preparation of nBSA and PSpBSA samples for aggregation experiments:** nBSA was dissolved in 10 mM phosphate buffer (PB) of pH 7.0 to prepare a 1.2 mM stock solution and stored at 4°C. Similarly, a 0.8 mM PSpBSA stock solution was prepared, but in Milli-Q water and stored at 4°C. Protein concentrations were determined by measuring the absorbance of tryptophan at 280 nm with a Thermo Scientific NanoDrop UV-visible spectrophotometer. The molar extinction coefficient of BSA at 280 nm is 43,824 $M^{-1} \cdot cm^{-1}$.[47] For the aggregation experiments, the stock protein solutions were diluted using a 50 mM PBS buffer (pH 7.0; 50 mM NaCl), containing a final protein concentration of 100, 150 or 200 µM. The protein samples were placed in glass vials and then heated in an oil bath at 65 °C for 2 hours, and then quenched on ice to store the samples.

**Thioflavin-T fluorescence assay**: The aggregation of nBSA in the amyloid structures was measured by monitoring the Thioflavin-T (ThT) fluorescence as a function of the time the protein solutions were kept at 65°C. To this end, we extracted small aliquots (50 µL) from the heated 100 µM protein samples in 10 mM PB of pH 7.0 with increasing incubation times. The extracted aliquots were cooled on ice to stop the denaturation process, and then diluted 10-fold with 10 mM, pH 7.4 PB, containing ThT, such that the final ThT concentration was 18 µM ThT (following a standard procedure reported in literature[46]). In the ThT-fluorescence intensity measurements, we used the standard excitation wavelength $\lambda_{ex}$ = 450 nm used for detecting amyloid fibril formation, while the emission was monitored in a wavelength range of 465-650 nm, using a Cary Eclipse fluorescence spectrophotometer. The slit width was 5 nm in both excitation and emission measurements, and all fluorescence measurements were performed at 25°C. All measurements were repeated three times to get the standard error of the measurement.

**Fourier transform infrared (FTIR) spectroscopy:** FTIR analysis of the samples was performed using a Thermo Fisher Scientific Nicolet iS10 FTIR spectrometer. A droplet of 10 µL of protein solution was deposited on a transparent KBr pellet and immediately dried using a mild nitrogen flow. The pellet was then placed in a transmission holder and the IR spectra were acquired in a range of 1500-1700 cm$^{-1}$. The background measurement was performed following a similar procedure using pure buffer solution. The dried films of nBSA fibrils were sandwiched between two thin KBr pellets and then placed in the transmission holder for spectra acquisition. For each measurement, 32 scans were collected with a resolution of 2 cm$^{-1}$ in transmission mode and the corresponding buffer spectrum was subtracted for each measurement. The amide I region (1600-1700 cm$^{-1}$) of the FTIR data was deconvoluted with the help of a Fourier self-deconvolution (FSD) method (see SI methods for details). The area under each deconvoluted peak was used to estimate the secondary structure in terms of α-helix and β-sheet fractions of the protein.

**Particle sizing and microrheology using DLS:** A Malvern Zetasizer ZSP (633 nm HeNe laser) was used to perform both measurements of the size of the proteins and their aggregates as well as the change in the viscoelastic behavior of the protein samples as function of time. For particle sizing, the Zetasizer was operated in backscattering mode at a scattering angle of 173° and measurements were performed on the fresh and heat-set solutions of nBSA and PSpBSA. In the DLS-based microrheology, we added 230 nm large, PEG-coated polystyrene spherical particles (Cambridge Bespoke Colloids, UK) to the protein solutions, which served as tracers. To ensure detection of single-scattering events only, we operated our setup in backscattering mode at a scattering angle of 173°, which we refer to as non-invasive backscatter (NIBS) detection mode.[48] The volume fraction of probe-particles used was 0.03% in all measurements, ensuring that the dominant scattering (ca. 95%) was due to the colloids and not due to the proteins or their aggregates. In all DLS experiments, we used Malvern ZEN0040 disposable cuvettes (40 μL) and all measurements were performed at 25°C. The measured intensity autocorrelation functions $g^{(2)}(q, t)$ were converted into intermediate scattering functions $g^{(1)}(q, t)$ using the Siegert relation, $g^{(2)}(q, t) = B(1+\alpha|g^{(1)}(q, t)|^2)$, where B is a baseline that is experimentally determined and $\alpha \approx 1$ is detector-dependent. For spherical particles of radius $R$, the intermediate scattering function writes as $g^{(1)}(q, t) = \exp(-q^2Dt)$, where $D$ is the translational particle diffusion coefficient that is defined through the Stokes-Einstein relation $D = k_B T(6\pi\eta R)^{-1}$ with $T$ being the temperature, $\eta$ the viscosity of the solvent and $k_B$ the Boltzmann constant. The diffusion coefficient of the probe-particles in 3 dimensions is also related to the mean-squared displacement (MSD) through the Einstein relation

MSD = $\langle\Delta r(t)^2\rangle$ = $6Dt$. Hence, we extracted the MSDs from our correlation functions $g^{(1)}(q, t)$ = exp($-q^2\langle\Delta r(t)^2\rangle/6$) using an in-house developed MATLAB routine.[48] The MSDs were then Fourier transformed to obtain the elastic, $G'(\omega)$, and loss, $G''(\omega)$, moduli of the protein solutions using other, in-house developed MATLAB routines[48]. This transform is based on the Generalized Stokes-Einstein Relation $\langle(\Delta\tilde{r}(\omega))^2\rangle \propto k_B T(6\pi R G^*(\omega))^{-1}$, where the complex shear modulus is defined as $G^*(\omega) = G'(\omega) + iG''(\omega)$.

**Film formation with nBSA fibrils:** In a typical experiment, 200 μL of nBSA solution (100 μM) in PBS (10 mM, pH = 7.0) containing 50 mM NaCl solution was poured into a 600 μL plastic Eppendorf tube (Cliklok microcentrifuge tubes; yellow, T330-6Y) and sealed to minimize evaporation. The sealed tubes were then incubated at 65 °C without agitation in a silicon oil bath for a maximum of 2 days. In the ThT measurements, we extracted a probe after 2 h and quenched it on ice, while FTIR spectroscopy was performed on the actual film that formed after 2 days.

**Visualization of nBSA:** We used the freely available Visual Molecular Dynamics (VMD) software to illustrate the secondary structure of native BSA, using the protein sequence of nBSA that is available on the Protein Databank (pdb: id 3v03).

**RESULTS**

**nBSA and bioconjugates of BSA under aggregation conditions:** The aggregation behavior of different concentrations of nBSA and its bioconjugate in PBS was investigated by monitoring the changes in the Thioflavin-T (ThT) fluorescence intensity. ThT is a well-known fluorescent amyloid marker that exhibits an increase in fluorescence intensity with an emission peak at 485 nm in the presence of β-sheets.[49] Figure 1A shows the VMD representation of the typical α-helix dominated secondary folded structure of nBSA monomers in the aqueous solution. After incubating 100 µM nBSA samples at 65°C in an oil bath for 2 hours, a maximum in the broad absorption peak was reached. The ThT-fluorescence spectra were measured at 25°C and are shown in Fig. 1C: the black dashed line is the averaged spectrum obtained for pure PBS containing ThT, while the solid black line was measured in the presence of freshly prepared nBSA (100 µM). The blue curve shows the averaged spectrum of the same sample measured after the aggregation conditions were applied for 2 h and then cooled on ice to stop further aggregation. More detailed, time-resolved ThT-fluorescence measurements for 100, 150 and 200 µM nBSA solutions with the same protein-to-ThT mass ratio are shown in the supporting information (Fig. S1). These measurements demonstrate that a maximum in the ThT-fluorescence peak was reached after around 90 min for the heated 100 µM nBSA solutions and after 2 h for the higher concentrations. These findings suggest that all proteins were converted into amyloid filaments. The VMD pictorial representation of the polymer surfactant conjugated PSpBSA monomer is shown in Fig. 1D. Figures 1E and 1F show the pictorial and ThT-fluorescence spectra of the PSpBSA monomers after subjecting their solutions to the same heat treatment and solvent conditions: these show no indication of amyloid filament formation.

**DLS study of nBSA and PSpBSA under aggregation conditions:** The growth of the filamentous aggregates formed by nBSA after heating the samples for up to 120 h was monitored with DLS at room temperature and compared to the intensity correlation functions measured for the fresh, non-heated sample prepared in identical solvent conditions (Fig. 2A). In parallel, we also studied freshly prepared PSpBSA solutions that were subsequently subjected to a similar heating protocol. Figure 2C displays the normalized translational diffusion coefficient $D/D_0$ as function of time, where $D_0$ and $D$ represent the translational diffusion coefficients of the fresh and heat-set protein solutions, respectively.

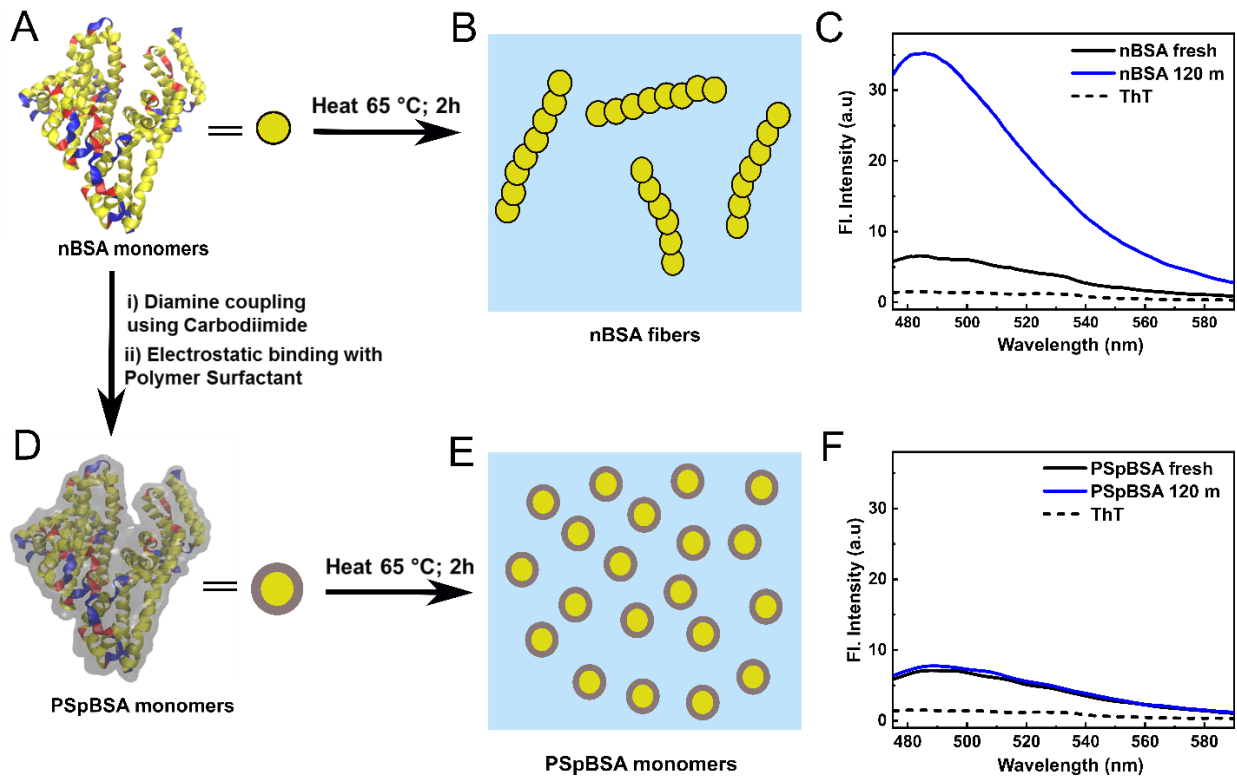

**Figure 1.** (A) VMD graphical representation of a nBSA monomer (α-helixes are yellow, β-sheet are blue and the red stands for random coil), sketched as yellow disk. (B) Schematic illustration of how nBSA monomers aggregate to form filaments. (C) ThT-emission spectra taken from a fresh 100 μM nBSA solution (pH 7.0, 50 mM NaCl, 25°C) and the same solution heated for 2 h at 65 °C. For reference, the dashed line was measured for pure PB containing the same final ThT concentration as the protein samples. (D) Graphical representation of polymer surfactant conjugated PSpBSA. The grey layer enveloping the protein represents the polymer surfactant (PS) layer coating each monomer (here represented as grey-rimmed yellow disks). (E) Schematic representation of PSpBSA monomers after subjecting them to aggregation conditions. (F) ThT-

fluorescence spectra of the fresh and heat-set PSpBSA solutions. As in (C), the dashed line represents the fluorescence activity of ThT in pure PBS. Again, the same protein and ThT concentrations were used (see Experimental Section).

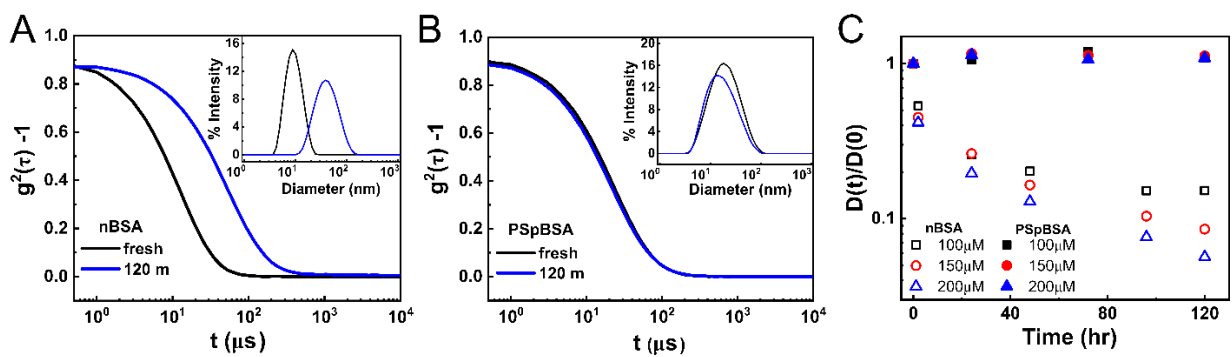

**Figure 2.** DLS measurements of 100 µM nBSA and 100 µM PSpBSA solutions. (A, B) show the directly measured intensity autocorrelation function of the fresh and heat-set nBSA and PSpBSA solutions under aggregation conditions. The insets represent the corresponding intensity distributions as functions of measured particle sizes. (C) Normalized diffusion coefficients of nBSA (open symbols) and PSpBSA (filled symbols) in solution as function of the time they were kept at 65°C (PBS at pH 7.0, containing 50 mM added NaCl). Here black, red and blue symbols represent the 100, 150 and 200 µM nBSA and PSpBSA concentrations, respectively.

**Secondary structure analysis by FTIR spectroscopy:** To understand the secondary structure of the protein, we performed FTIR studies at different time points during the aggregation. Figure 3 shows the Fourier self-deconvoluted FTIR spectra (see SI methods for details) taken in the amide I region (1600-1700 cm$^{-1}$) of the fresh nBSA and PSpBSA solutions and after subjecting them for 24 h to our aggregation conditions.

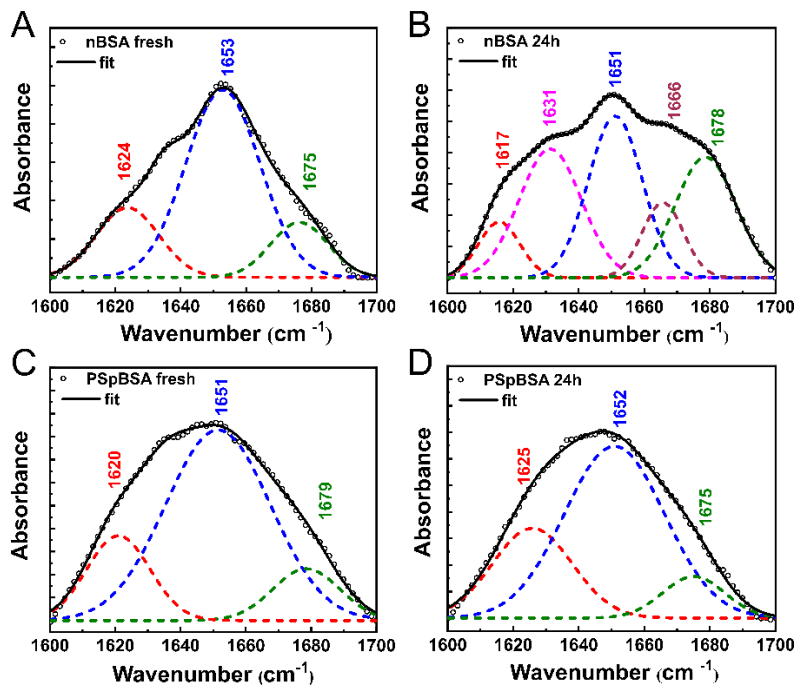

**Figure 3.** Secondary structural changes of nBSA and PSpBSA monitored by FTIR spectra, in the region corresponding to the amide I band (1600-1700 cm$^{-1}$), (A) nBSA fresh, (B) nBSA fibrils after 24 h heating at 65°C, (C) PSpBSA fresh, and (D) PSpBSA after 24 h heating at 65 °C. The open symbols represent the experimental values, whereas black solid line shows the fits to the experimental data. The dashed lines show the different components, obtained after Fourier self-deconvolution (see Table 1).

**DLS microrheology study of nBSA aggregation:** We have studied the viscoelastic properties of the nBSA solutions under aggregation conditions using DLS-based microrheology. This was achieved by measuring the time-dependent correlation function of the light backscattered from 230 nm large, spherical probe-particles dispersed in the protein sample. From the measured and normalized intensity autocorrelation functions, we first extracted the intermediate correlation function $g^{(1)}(q, t) = \exp(-q^2 Dt) = \exp(-q^2 \langle \Delta r(t)^2 \rangle / 6)$, which relates the particles' translational diffusion coefficients to the MSDs that describe their thermal motion in the viscoelastic medium.[50] In Figure 4A, we show the $g^{(1)}(q, t)$ curves for a freshly prepared and heat-set nBSA samples measured as function of heating times of up to 7 days for the three different concentrations of nBSA (100, 150 and 200 µM); the corresponding MSDs are shown in Figure 4B. For comparison, we measured $g^{(1)}(q, t)$ for tracer particles dispersed in pure water (dashed blue curve in Figure 4). It showed the expected single relaxation time, $\tau = (q^2 D)^{-1}$ that is inherent to the particles diffusing freely in a Newtonian fluid. The 100 µM and 150 µM nBSA solutions showed almost identical scattering curves as those for the particles in pure water, with very little change even when heated for 7 days. Only the 200 µM nBSA solution showed a measurable change with increased relaxation times, which manifested themselves in a sub-diffusive behavior visible in the corresponding MSD curves. Hence, we show in Figure 5 the elastic and viscous moduli ($G'(\omega)$ and $G''(\omega)$), obtained from the Fourier transforms of the MSD curves for the 200 µM nBSA sample.

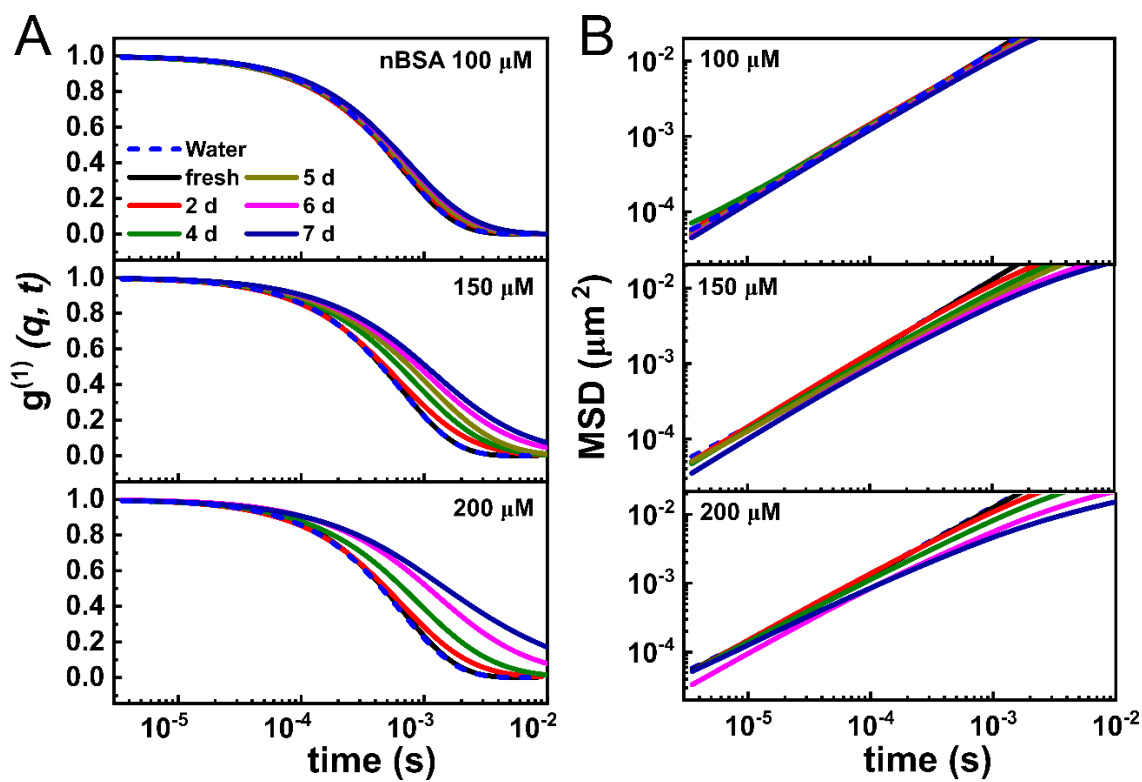

**Figure 4.** (A) Time-dependent intensity autocorrelation curves, measured for the 100, 150 and 200 μM nBSA solutions under the same aggregation conditions mentioned before, containing 0.03 (vol/vol%) 230 nm PEG-coated polystyrene latex tracer particles. The curves were measured starting from fresh solutions (presented as black curves) up to samples heated at 65°C for seven days (navy blue curves). (B) Corresponding MSD data. For comparison, we present calculated MSD curves for the same 230 nm colloids in pure water (blue dashed lines).

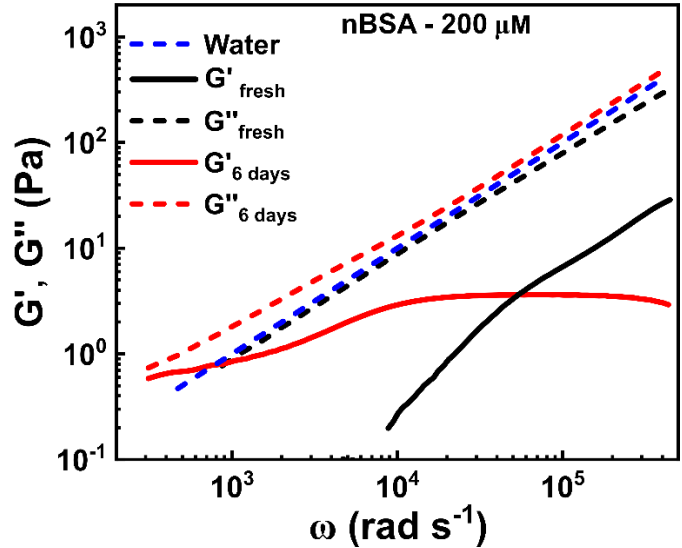

**Figure 5.** Time evolution of the elastic ($G'(\omega)$) and viscous ($G''(\omega)$) moduli as functions of frequency, extracted from the MSD curves in Figure 4B of 200 μM nBSA in solution under aggregation conditions.

**Film formation from nBSA fibrils:** In Figure 6, we present the scheme for film formation in a 200 μM nBSA solution incubated at 65 °C in a plastic Eppendorf tube up to 2 days (see Method for details) and the related fluorescence and FTIR studies. In the first few hours of heating the sample at 65°C under aggregation conditions, no visible aggregates were detected in the sample. However, we performed time-resolved ThT-fluorescence spectra on the protein solutions. The maximum of the measured fluorescence peak at 485 nm was plotted as a function of heating time in Figure 6D. It shows a fast, initial growth that slows down considerably after about 40 minutes, similar to the ThT fluorescent behavior measured for 200 μM nBSA solutions heated in glass cuvettes, demonstrating the formation of amyloid-type filaments. A clearly visible film is only detected after heating the sample in an Eppendorf tube for at least one day. A photograph of such a film, grown for 2 days is shown in Figure 6C. We performed FTIR spectroscopy on this film – the results are shown in Figure 6E. The spectrum shows strong peaks corresponding to β-sheet structures as well as additional peaks, which we will discuss further in the next section.

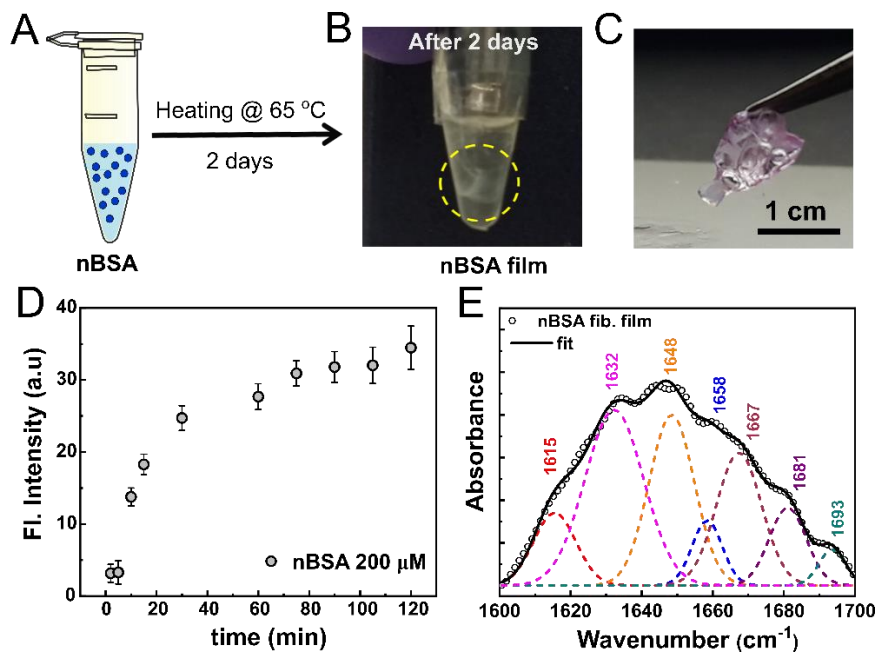

**Figure 6.** (A, B) Scheme of nBSA film formation from fibrils. (A) shows the nBSA monomers in an Eppendorf tube, and the formed film after 2 days of incubation is shown in B. (C) Pictorial representation of the self-assembled nBSA fibril film. The film is stained with Nile red dye for clarity. (D) Thioflavin-T (ThT) fluorescence intensity curve for the initial 120 minutes of nBSA film formation. (E) Fourier self-deconvoluted FTIR spectrum of the amide I region of the nBSA film, indicating its predominant β-sheet structure.

**DISCUSSIONS**

Recently, conjugation of proteins with oppositely charged PEG-based polymer surfactants has been explored.[38–40] Interestingly, these conjugated globular proteins retain a near-native secondary folding structure, while the same conjugation of the intrinsically disordered protein α-synuclein (showing no secondary structure) promotes ordering and the formation of highly stable α-helices in a crowded liquid.[41,51] These findings motivated us to understand the stability of the preserved secondary structure in bioconjugated proteins in solution under protein aggregation conditions. We have selected the globular carrier protein BSA[52] as model protein to study the aggregation behavior of its native and its stoichiometric bioconjugate under neutral pH conditions. Several studies explored the aggregation behavior of native BSA as function of pH, concentration and temperature, and showed that these aggregates have amyloid-like properties. [42–45] Here, we studied the aggregation kinetics of three different nBSA and PSpBSA concentrations (100, 150 and 200 µM) that were incubated at 65 °C in an oil bath. To allow for equilibrium growth of the aggregates, we avoided agitation or stirring of the samples. All solutions were prepared in PBS with a final pH 7 containing 50 mM NaCl in glass containers. Holm et al.[44] reported that nBSA denatures and forms amyloid fibrils around 65 °C in pH 7.4 and conducted time-resolved ThT-fluorescence spectroscopy between 60 °C and 75 °C. Their findings showed that the aggregation process was too fast to follow at higher temperatures, while at 60°C no significant amyloid formation was observed. Therefore, we decided to study the aggregation dynamics of nBSA and PSpBSA with time-resolved ThT-fluorescence experiments at 65 °C (Figure 1). Further, we used for all three nBSA concentrations the same 1:2 protein-to-ThT molecule ratio, which has been reported to be the optimal ratio, leading to the strongest excitation at around 485 nm, with one ThT molecule fitting into the right-sized cavity formed between two β-sheets. Similarly to Holm et al.,[44] we observed an initial sharp increase in the ThT-fluorescence intensity peak at 485 nm in all nBSA concentrations under aggregation conditions, which reached a plateau after around 60 minutes and showed no further significant change when incubated for another hour (Figure 1C).

The complete time evolution of the ThT spectra for 100, 150 and 200 µM nBSA solutions is shown in Figure S1. In agreement with earlier reports, the relatively fast plateauing in the peak of the ThT-fluorescence spectra suggests that within the first 60 minutes of heating all nBSA monomers have changed their native secondary structure, rich in α-helix, into a different configuration that enables them to aggregate into amyloyd-like filaments. Interestingly, time-resolved ThT-fluorescence studies in 100 µM PSpBSA solutions under aggregation conditions showed no significant increase in the fluorescence-peak intensity with respect to the fresh PSpBSA solution (see blue and black curves in Figure 1F). This indicates that unlike nBSA, the bioconjugate PSpBSA does not change its secondary structure under the influence of denaturation temperatures and thus does not aggregate under the aggregation conditions chosen here, even after 120 minutes. This absence of aggregation (Figures S2 and S3) and thus the heat-stability of PSpBSA is further confirmed by the relatively little change observed in the FTIR spectra and control measurements using Circular Dichroism (CD) spectroscopy, obtained for the fresh and heat-treated samples (Fig. 3 and Fig. S10).

To better understand the fibril formation of nBSA and the thermal stability of PSpBSA solutions, we performed various control experiments on the short-term and long-term aggregation behavior of the native and polymer surfactant conjugated protein solutions employing DLS measurements. We first investigated the growth of the nBSA fibrils under prolonged heating (aggregation). For reference, we first performed DLS measurements on the fresh protein samples that have not been exposed to the aggregation conditions yet (Figure 2). The intensity autocorrelation functions, $g^{(2)}(q, t)$, of both fresh nBSA and PSpBSA solutions show a single relaxation time, characteristic of monodisperse quasi-spherical particles with an average hydrodynamic radius $R_h$ = 4.5 nm for nBSA, and $R_h$ = 9 nm for the polymer surfactant coated PSpBSA under the present ionic strength of the solvent. The latter is expected to be slightly larger due to the coating with the polymer surfactant. These values appear somewhat overestimated; however, here we give the average values of the intensity evaluation of the DLS spectra. When plotting the number density as function of size of the scattering object, we observed a smaller value that is close to literature data, coexisting with a small fraction of larger particles, which we attribute to the presence of small oligomers that can appear through the polymer-surfactant coating process. It should also be noted that the DLS measurements were done with polarized laser light. In this situation, we measure purely the translational diffusion coefficient $D = k_B T (6\pi \eta R_h)^{-1}$. The measured $g^{(2)}(q, t)$ of very dilute rods can still be fitted using the Generalized Stokes-Einstein Relation; however, the measured hydrodynamic radius must be replaced by the radius of gyration $R_G$, reflecting the translational motion of the rods' centre of mass. Doi and Edwards[53] give an expression for the new translational diffusion coefficient, which depends on the length $L$ and thickness $d$ of the rods as $D_G = k_B T \ln(L/d)/(3\pi \eta L)$.

In Figure 2C, we show the measured translational diffusion coefficients $D(t) = D(t_{heat})$, normalized by the diffusion coefficient of the fresh nBSA and PSpBSA monomers $D_0 = D(t_{heat} = 0)$, respectively, as function of heating times $t_{heat}$ at 65 °C for up to 120 h. The individual scattering spectra for 100, 150 and 200 µM nBSA are given in the supporting Figures S4-S6, and those for 100 µM and 200 µM PSpBSA - in Figures S7 and S8. For the heat-set nBSA samples, the ratio $D(t)/D_0 = D_G(t)/D_0 = \ln(L/d)/(L/d)$ clearly decreases over the heating times we measured, reflecting the slow growth of the protein fibrils. We also observe a faster growth rate for larger nBSA concentrations. All concentrations show a slowing down in growth with time and we observe two quasi-linear regions. This is particularly visible in the 120 h heating study of the 100 µM nBSA sample, where the ratio starts to plateau after about 100 h of heating. The plateau occurs at $D(t)/D_0$

= 0.15, which suggests an average length of the fibrils of about 180 nm, assuming a fibril-diameter of about 9 nm corresponding to that of fresh nBSA monomers. We hypothesize that the slow growth of the fibrils follows a special case of reaction-limited aggregation mechanisms,[54,55] observed in many particulate systems. But unlike systems with dendritic growth, the protein fibrils can only grow linearly, as was observed in TEM images of heat-set nBSA samples.[44] Oozawa and Kasai[56] introduced a theoretical model explaining linear growth in proteins and di Michele et al.[57] reported simulation studies on self-catalytic, linear growth in proteins. Hence, we argue that within the first two hours of heating all nBSA monomers are converted into dimers or trimers, showing increased ThT-fluorescence, but the subsequent growth into long filaments occurs by end-to-end fusion of these short dimers to increasingly larger filaments. This end-to-end growth of smaller linear, semi-flexible rods will be reaction-limited, and it seems to be dependent on a temperature-dependent activation barrier. This hypothesis is supported by the scattering intensity-to-size plots for the two larger nBSA concentrations that show a remaining fraction of short dimers or trimers coexisting with the much larger fraction of long filaments after 120 h of heating (Figure S6).

In order to substantiate our hypothesis, we performed control measurements, in which we stopped the heat-induced conformational change in the nBSA secondary structure after 20 minutes, such that only a fraction can overcome the activation barrier for two 'seed' clusters to form and rearrange into amyloid structures. The ThT-fluorescence spectra in Figure S11 clearly show that stopping the heating of a 100 µM nBSA solution after 20 minutes halts further growth of the fluorescence peak. Indeed, DLS measurements of the 20-minute heated sample show a bimodal distribution with more nBSA monomers coexisting first with dimers and then with slightly larger rods. The DLS measurements of a 200 µM nBSA solution, which was heated only for 2 h and then kept at room temperature for days, showed no further growth of the initially formed small clusters (Fig. S11D), which is in contrast to the fibrillar growth observed for continuous heating (Fig. 2C).

While ThT-binding fluorescence data provides indirect information of the structure of heat-treated protein solutions, FTIR experiments (Figure 3A) give information on the secondary structure of proteins in terms of α-helix and β-sheet contributions. Fresh nBSA has a predominant peak at 1653 cm$^{-1}$ in the amide I region, which is characteristic of a protein, rich in α-helix. Fourier self-deconvolution (see SI for details) of the spectra gives us 65% α-helical content for our fresh nBSA samples (Table 1), which reduces to 39% after 24 h. The reduction in α-helical content is accompanied by an increase of β-sheet content in the BSA fibrils (Figure 3B and Table 1). This resembled FTIR studies on the sol-to-gel transition in regenerated silk fibroin (RSF) protein solutions.[58,59] Such a transition was associated with the conversion of random coil to β-sheet upon increasing concentration, mechanically vortexing or heating the RSF solution. The formation of amyloid-like structures during gelation was not observed in this case. However, recently there have been reports[60] of aggregation of RSF, resulting in the formation of amyloid fibrils, having diameter of 3.2 nm under certain conditions.

**Table 1.** Secondary structural content analysis from the deconvoluted FTIR spectra (Figure 3) of protein solutions.

| Sample | Peak (cm$^{-1}$) | Secondary structure | Content (%) |
|---|---|---|---|
| **nBSA fresh** | 1624 | β-sheet | 19 |
| | 1653 | α-helix | 65 |
| | 1675 | antiparallel β-sheet | 14 |
| **nBSA** | 1617 | β-sheet | 8 |
| **24 h** | 1631 | β-sheet | 27 |
| | 1652 | α-helix | 29 |
| | 1666 | β-turns | 10 |
| | 1678 | antiparallel β-sheet | 25 |
| **PSpBSA fresh** | 1620 | β-sheet | 18 |
| | 1651 | α-helix | 66 |
| | 1679 | antiparallel β-sheet | 12 |
| **PSpBSA 24 h** | 1625 | β-sheet | 26 |
| | 1652 | α-helix | 64 |
| | 1675 | antiparallel β-sheet | 10 |

Comparison of the FTIR spectra of fresh nBSA and PSpBSA solutions shows that both are very similar, with a main peak at around $1652 \pm 2$ cm$^{-1}$, indicating a secondary structure, rich in α-helix, and only a small discrepancy between the β-sheet and antiparallel β-sheet contributions. Upon subjecting PSpBSA to aggregation conditions, we observe very small changes in the FTIR spectrum confirming that heating at around the native protein's denaturation temperature does not change the bioconjugated counterpart (Figure 3D and Table 1). This is even true when heating the PSpBSA samples at 65 °C for many days or even heating them up to 80 °C for several hours (Figure S9, Figure S10, Tables S2 and S3). Hence, our experimental results infer that stoichiometric conjugation of PEG-based polymer-surfactant renders BSA remarkably inert and prevents it from forming amyloid aggregates. This opens up an important pathway to develop drugs, foods and household products with much-improved shelf-life and efficacy.

The effect of fibril formation will also be expressed in the viscoelastic properties of our relatively dilute protein solutions. The fresh solutions will have viscosities similar to those of the continuous background solvent. The Einstein relation for the viscosity of a dilute suspension of spherical colloids is $\eta = \eta_w (1 + 2.5\phi + ...)$, where $\eta_w$ is the viscosity of water and $\phi$ is the volume fraction of the colloids. We can assume that nBSA can be approximated as a spherical colloid with a diameter of 9 nm, which means $\phi$ is roughly 2%. This leads to a change in the average viscosity of less than a percent in the present case. Indeed, the DLS-based microrheology measurements of the three different, fresh nBSA concentrations show an almost identical MSD and thus diffusivity as those for pure water (Figure 4B). However, when the nBSA molecules upon heating aggregate into sufficiently long fibrils, we can expect an overall change in viscosity and an onset of some elasticity, if the rods start to touch each other, thus giving rise to excluded-volume interactions. Our microrheology experiments provide understanding of the equilibrium elastic and viscous moduli of the heat-treated nBSA solutions over a very large frequency range, reaching MHz, which conventional rheology cannot provide. In particular, we can expect contributions to the system's elasticity at higher frequencies, while at lower frequencies it is the viscous part that will dominate. The decay of $g^{(1)}(q, t)$ is related to the time-evolution of the probe-particle motion, allowing for the MSD of the particle to be measured. The MSD in turn is related to the viscoelastic modulus via the Generalized Stokes–Einstein Relation.[61,62]

The decay of $g^{(1)}(q, t)$ slows down over the time interval from fresh to 7 days under continuous heating at 65 °C, indicating an increased viscosity of the medium (Figure 4A). This slowing down in relaxation time becomes more pronounced as the protein concentration increases. The concentration-dependence is particularly noticeable for the longer timescales in the MSD curves (Figure 4B), when the long filaments have a chance to interact via excluded-volume interactions. As expected, all short-time responses coincide with the free diffusivity of the individual proteins or aggregates. Figure 4B indicates that at concentration 100 μM nBSA, MSD curves depend linearly on the lag time $\tau$ over the time period from fresh solution to 7 days, confirming the $\langle \Delta r(t)^2 \rangle \propto \tau$ relation for Newtonian fluids and thus proving that nBSA behaves like a fluid at low concentration under aggregation conditions. Heat-set 200 µM nBSA solutions show initially (up to 2 days) similar MSD curves. However, they have a slightly lower exponent, indicating sub-diffusive motion of the local nBSA fibrils. At longer relaxation times, corresponding to the intermediate $\omega$, the MSD curves reach a plateau. In Figure 5, the dashed blue line represents the loss modulus of water, which is a Newtonian liquid and thus with a loss modulus changing linearly with viscosity: $G''(\omega) = \eta \, \omega$. The fresh 200 µM nBSA solution (black dashed curve) shows similar viscosity. As Newtonian fluids have no elasticity, we do not show $G'(\omega)$ for water, but the fresh nBSA solution does show some elasticity at very high frequencies, which significantly increases as the time progresses, while $G''(\omega)$ retains the linear frequency behavior of the fresh sample. After 6 days of aggregation conditions, $G'(\omega)$ reaches a plateau value of $\sim$ 3.58 Pa (Figure 5) in the intermediate time range, corresponding to a possible entanglement length $\xi \sim 105$ nm, assuming the scaling behavior of the bulk modulus to be $G_{bulk} = k_B T / \xi^3$.[53] Similarly, we measured plateau values of $\sim$ 1.2 Pa and $\sim$ 3.0 Pa (Figure S13) for 100 µM and 150 µM concentrations of nBSA fibrils, showing the right trend that lower concentrations will deliver shorter fibrils for the same heating time. Note that these values are only approximate and have an error of 20-30%.

While observing standard fibril formation when heat-setting nBSA solutions, we discovered a completely different aggregation behavior when incubating the nBSA solutions (here using 200 µM) under our aggregation conditions for 2 days in plastic Eppendorf tubes (Fig. 6): a stable film of thickness ~1.6 µm (Figure S14) formed. To understand the nature of the film, we performed ThT-fluorescence experiments in the early stage of film formation. Within the initial 120 minutes of heating, we found a sharp increase in the ThT-fluorescence intensity (Fig. 6D), confirming the formation of amyloid aggregates inside the Eppendorf tubes. Also, experiments with 100 μM and 150 μM nBSA solutions under the same incubation conditions resulted in protein films with similar ThT-fluorescence findings (Figure S15 and Figure S16). Incubating bioconjugated PSpBSA under the aggregation conditions in an Eppendorf tube, on the other hand, showed no change even after 5 days, again showing high stability against heat-induced aggregation. Previous reports on Aβ$_{1-40}$ protein solutions state that amyloid fibrils have a tendency to get adsorbed onto the hydrophobic surfaces of Eppendorf tubes.[63] We believe that similarly, nBSA fibrils in our experiment were adsorbed to the surface of the hydrophobic plastic tube and then self-assembled to form protein films. However, placing the films in fresh buffer solution and sonicating them did not re-disperse the filaments. Hence, the aggregation process in the Eppendorf tubes must be accompanied by some crosslinking mechanism that still needs to be studied. Interestingly, comparing the FTIR experiments performed on the protein films with the heat-set nBSA fibrils formed in glass cuvettes (Fig. 3B) showed that the films contain a 40% β-sheet secondary structural content and an additional peak at a wave number of 1693 cm$^{-1}$. This confirms the presence of an additional process occurring simultaneously (Figure 6E and Table S1), which may be similar to the aggregation processes observed in regenerated silk fibroin protein solutions.[58,59]

**CONCLUSIONS:**

To summarize, we demonstrate here the exceptional thermal stability of polymer-surfactant conjugated BSA under aggregation conditions. Our combined ThT-fluorescence, DLS and FTIR results give us a clear description of bioconjugated BSA under fibrillation conditions and its comparison with native BSA under similar aggregation conditions. Stability and protection against aggregation for the bioconjugates of BSA (PSpBSA) appear to correlate with the role of PEG-based polymer-surfactant surrounding the native protein. The polymer-surfactant layer surrounding the protein presumably hinders the interprotein contacts and thus prevents the aggregation of protein. DLS measurements also shed light onto the fibril formation of nBSA under varying heating conditions. Further, our microrheological measurements show that the native BSA fibrils form a network of semi-flexible polymers after prolonged aggregation times. But we also show that the nature of aggregation can depend on the environment, such as the container walls. Our study has a huge potential in finding alternative formulations compared to the commonly used osmolytes and detergents that rely on stable protein solutions.

## ASSOCIATED CONTENT

**Supporting Information** includes experimental details, Thioflavin-T fluorescence studies, FTIR, DLS data.

## AUTHOR INFORMATION

**Corresponding Author**

* Erika Eiser: ee247@cam.ac.uk

**Present Addresses**

†If an author's address is different than the one given in the affiliation line, this information may be included here.

**Author Contributions**

‡ A. M. and I. D. S. contributed equally to this work.

**Funding Sources**

Any funds used to support the research of the manuscript should be placed here (per journal style).

**ACKNOWLEDGMENT**

A. M. acknowledges Commonwealth Scholarship Commission (UK government) and the Department of Chemistry, IITB for financial support. I. D. S. and D. A. K. thank EPSRC for financial support. K. P. S. thanks IRCC and DBT India for funding through grant no. EMR/15IRCCSG0029. E. E. acknowledges the ETN-COLLDENSE (H2020-MCSA-ITN-2014, Grant No. 642774).